\documentclass[%
 reprint, amsmath,amssymb, aps,showpacs]{revtex4-1}
\usepackage{graphicx}
\usepackage{epsfig}
\begin{document}


\title{Floquet dynamics in multi-Weyl semimetals}
\author{Amit Gupta}

\affiliation{%
Theoretical Sciences Unit, Jawaharlal Nehru Centre for Advanced Scientific Research, Bangalore-560064, India
  \\}%
  
\date{\today}

\begin{abstract}
Multi-Weyl semimetals(m-WSMs) are characterized by anisotropic non-linear energy dispersion along 2-D plane and a linear dispersion in orthogonal direction. They have topological charge J and are realized when two or multiple Weyl nodes with nonzero net monopole charge bring together onto a high-symmetry point. The model Hamiltonians of such systems show various phases such as trivial normal insulators (NIs), WSMs and quantum anomalous Hall(QAH) states depending on the model parameters. We study a periodic driving of such Hamiltonians in trivial NI phase by circularly polarized light and shows this phase can be tunable to Weyl semimetals phase. This transition can be accomplished with large anomalous Hall conductivity which could be observed in transport measurements or pump-probe angle resolved photoemission spectroscopy experiments.
\end{abstract}

\maketitle


\section{Introduction}
Weyl semimetals(WSMs) are three dimensional analogue of graphene where the low energy Hamiltonian has linearly dispersing excitations [which obey the 3D Weyl equation] from degenerate band touching points referred to as Weyl points or nodes in all the three $ \textbf{k} $ directions. This Weyl points are protected by the monopole charge\cite{vafek14}. As a consequence, Weyl semi-metals (WSMs) host topologically protected surface states in the form of open Fermi arcs terminating at the projection of bulk Weyl points of opposite chirality. Another important physical consequences in WSMs are exotic transport phenomena such as a large negative magnetoresistance due to chiral anomaly \cite{volovik86, volovik09, son13}. In recent angle-resolved photoemission spectroscopy(ARPES) and scanning tunneling microscopy(STM) experiments, several materials such as Cd$_3$As$_2$ \cite{neupane14,jeon14,liu14,borisenko14,he14,li15,moll15}, Na$_3$Bi \cite{liu14a}, NbAs\cite{xu15a},  TaP\cite{xu15, xu16}, ZrTe$_5$ \cite{chen15, li16}  and TaAs \cite{lv15,lv15a,lv15b,huang15a,xu15b,yang15,inoue16} have been identified as WSMs. Several other efforts have been made on the realization of WSMs in artificial systems such as photonic crystals \cite{chen15a,zhang15,dubcek15,lu15,lu16}.\\

In addition to above isotropic linearly dispersive WSMs, a new three-dimensional topological semimetal has been proposed in materials with $C_ {4,6}$ point-group symmetries \cite{xu11,fang12,huang16}. This are referred as multi-WSMs. This could be achieved by merging of Weyl points of the same chirality. e.g. the double(triple)-Weyl semimetals can be achieved when two Weyl points of chirality $ + 1,+2$ or $(-1,-2) $ merge together and merging points have quadratic(cubic) dispersions along $k_{x,y}$ plane and linear dispersion along $k_z$ direction. The double-Weyl point possesses a monopole(anti-monopole) charge of +2 (-2) and the double-Weyl semimetal shows double-Fermi arcs on the surface Brillouin zone(BZ) \cite{xu11,fang12,huang16}. Similarly, triple-Weyl point possesses a monopole(anti-monopole) charge of +3 (-3) and it shows triple-Fermi arcs on the surface BZ. The double-Weyl points are protected by $C_4$ or $C_6$ rotation symmetry and its half-Metallicity has been realized in the three-dimensional semimetal $\mathrm{HgCr_2Se_4}$ in the ferromagnetic phase, with a pair of double-Weyl points along the z-direction \cite{guan15}. In $\mathrm{SrSi_2}$, another WSM, where Weyl points with opposite charges are located at different energies due to the absence of mirror symmetry \cite{huang16}. The experimental realization of triple-WSMs has not been reported so far. However, it is  impossible to realize Weyl points with monopole charge larger than three from an underlying lattice model Hamiltonians \cite{xu11,fang12,huang16}.\\


Our paper is organized as follows. In Sec. II, we study the periodic driving of multi-WSMs in trivial insulating phase by circularly polarized light along z-direction. The periodic driving of monopole charge J=1  has been discussed in the literature \cite{Awadhesh15, Chan16, Nagaosa16}. Here, we generalize it for higher monopole charge J. The consequence of such periodic driving is to tune trivial NI phase to WSM phase. This transition depends on the polarization of the laser field $\xi(\pm 1)$. In Sec. III, we discuss the main consequence of this transition to WSM phase is accomplished with large anomalous Hall conductivity which can be controlled by intensity and frequency of the laser light. Our predictions could be observed in transport measurements or pump-probe angle resolved photoemission spectroscopy. In Sec. IV, we present a brief conclusions and discussion of our results.


\section{Model Hamiltonian}
\label{mod_ham}
The non-interacting low energy effective Hamiltonian for a single multi-Weyl semimetals is given by \cite{xu11,fang12,huang16,guan15, Yang14,ahn16, ahna16},
\begin{equation}
\label{eq:ham}
\mathcal{H}_J=\alpha_J[( \hat{k}_{-})^J \sigma_{+}+ (\hat{k}_{+})^J\sigma _{-}]+\chi v_z\hat{k}_z\sigma _z,
\end{equation}
where $\sigma_{\pm}=\frac{1}{2}(\sigma_{x}\pm i\sigma_{y})$ and $\hat{k}_{\pm}=\hat{k}_x\pm i \hat{k}_{y}$,  J represents monopole charge, $\chi =\pm 1 $ represents the chirality, $v_z$ is the effective velocity along $\hat{z}$ direction and  $\alpha_{J}$ is the material dependent parameter, e.g. $\alpha_1$ and $\alpha_2$ are the Fermi velocity $ v_F$ and inverse of the mass m respectively for the single and double WSMs respectively. The energy dispersions of Eq.(\ref{eq:ham}),
\begin{equation}
E_{s}(\mathbf{k})=s\sqrt{\alpha_n^2 (k_{\parallel})^{2n}+\left(\hbar k_z v_z \right)^2},
\end{equation}

\noindent where $s=\pm$ and $ k_{\parallel}=\sqrt{k_x^2+k_y^2} $ is the momentum along $\hat{x}$-$\hat{y}$ plane. Throughout this paper, we have assumed  $\hbar =1$. The density of states of multi WSMs depends on their monopole charge $g(\epsilon )\sim \epsilon^{2/J}$ \cite{ahn16}. \\

A simple lattice Hamiltonian for the single-WSMs is given by \cite{Delplace12, Turner13, Chen15, ahna16},
\begin{eqnarray}
\label{swsm}
\mathcal{H}_1 = t_x \sin k_x \sigma_x + t_y \sin k_y \sigma_y + M(\textbf{k})\sigma_{z},
\end{eqnarray}
with lattice spacing unity and $ M(\textbf{k}) = m_z -t_z \cos k_z + m_0[2- \cos k_x - \cos k_y] $. The $ t_{x,y,z} $, $ m_z $ and $ m_0 $ are model parameters and are positive numbers. Similarly, we have generalized form of the above lattice model in Eq. (\ref{swsm}) for J = 2 and J= 3 such that near the Weyl points the low energy Hamiltonians reduces to the  Eq.(\ref{eq:ham}) \cite{Jian15, ahna16}

\begin{eqnarray}
\label{dwsm}
\mathcal{H}_2& =& t_x [\cos k_x -\cos k_y]\sigma_x + t_y \sin k_x \sin k_y \sigma_y \nonumber\\
&+& M(\textbf{k})\sigma_{z} ,\\
\mathcal{H}_3& =& t_x \sin k_x[-\cos k_x +\cos k_y -2] \sigma_x  \nonumber\\
&+& t_y[-\cos k_y + 3\cos k_x-2]\sigma_y + M(\textbf{k})\sigma_{z},
\label{twsm}
\end{eqnarray}

The model Hamiltonians Eq.(\ref{swsm})-(\ref{twsm}) show various phases such as normal insulators (NIs), WSMs and quantum anomalous Hall(QAH) states depending on the model parameters  \cite{Jian15, ahna16}. Note that the model Hamiltonians Eq.(\ref{swsm})-(\ref{twsm}) have two Weyl points with opposite chiralities in WSM phase. However,in general, there can be multiple Weyl points in the Brillouin zone(BZ) with total chiralities sum zero.

\section{Periodic laser field driving}

We consider the continuum model of above lattice Hamiltonians. The continuum Hamiltonian show various phases such as trivial normal insulators (NIs), WSMs and quantum anomalous Hall(QAH) states depending on the model parameters. We focus near $ \textbf{k} = (0, 0, 0)$ where WSM phase and trivial insulating phases arise. We start with a periodic driving of continuum Hamiltonian near $ \textbf{k}= (0, 0, 0)$ in the trivial insulating state and effects of periodic driving leads to WSM phase. 

\subsection{Single-WSMs}

A low-energy Hamiltonian for Eq.(\ref{swsm}) around  $ \Gamma $ point in the Brillouin zone, considering terms up to quadratic in $k$, is of the form~\cite{Delplace12, Turner13, Chen15, Wang12, Wang13}

\begin{equation}
 \mathcal{H}= v_F(k_{x}\sigma_{x}+k_{y}\sigma_{y}) + M(\mathbf{k})\sigma_{z},
\end{equation}

\noindent where $v_F$ is effective isotropic velocity along $k_x, k_y$ directions and $M(\mathbf{k})\approx m_z-t_z + \frac{1}{2}t_z  k_z^2 + \frac{m_0}{2}(k_x^2+k_y^2)$ near BZ center. The energy eigenvalues are obtained as $E_{\pm}=\pm\sqrt{v_F^{2}(k_{x}^{2}+k_{y}^{2})+M(\mathbf{k})^{2}}$. The energy spectrum is gapped throughout the BZ for $m_z > t_z$ and thus it is trivial normal insulator(NI). The gap parameter $m_z$ can be tuned by laser light  which renormalized the parameter $m_z$ which leads to WSM phase i.e. $m_z < t_z$ . The condition $m_z = t_z$ is the topological phase transition between NI phase and WSM phase. Our study can easily be extended to QAH-WSM transition near $ \textbf{k}=(0,0,\pm \pi) $. We can approximate $M(\mathbf{k})\approx m_z+t_z - \frac{1}{2}t_z  k_z^2 + \frac{m_0}{2}(k_x^2+k_y^2)$ near $ \textbf{k}=(0,0,\pm \pi) $ . For  $m_z < -t_z$, the energy spectrum changes its sign in the BZ and generates a non-trivial insulating quantum anomalous Hall phase in contrast to the case of NI phase. The condition  $m_z = -t_z$ is the topological phase transition line between QAH phase and WSMs. This transition can also be tuned by periodic driving. However, we will discuss only the transition between NI and WSM phase.  \\

When the system is exposed to circularly-polarized off-resonant light of laser frequency $\omega$, the response of the electrons can be obtained by minimal coupling $\textbf{k}\rightarrow \textbf{k}+e\textbf{A}(t)$. The circularly polarized can be expressed as $\textbf{A}(t)=A_0( \cos \omega t,\xi \sin \omega t, 0)$. Here $\xi=\pm 1$ for right and left circularly polarized beams, respectively and $ A_0$ is the amplitude of the laser \cite{argument1}. 
The full Hamiltonian is time periodic and thus, it can be expanded as Fourier series
$\mathcal{H}(t,k)=\sum_{n}\mathcal{H}_{n}(k)e^{in\omega t}$ with Fourier components
\begin{eqnarray}
\mathcal{H}_{0}(p)&=&v_F(k_{x}\sigma_{x}+k_{y}\sigma_{y})+  M(\mathbf{k})\sigma_{z},\nonumber\\
\mathcal{H}_{\pm1}(k)&=& \frac{eA_{0}}{2}v_F(\sigma_{x}\mp i \xi \sigma_{y})+m_0 e A_0 (k_{x}\mp i \xi k_{y})\sigma_z,\nonumber\\
\mathcal{H}_{\pm2}(k)&=&0,
\end{eqnarray}
and $\mathcal{H}_{n}=0$ for $|n|>2$. Using the commutation relations for the Pauli matrices, we find that the commutator has the form, $[H_{-1},H_{+1}]= 2\xi e^{2}A_0^{2}v_{F}m_0(k_{x}\sigma_{x} + k_{y}\sigma_{y} ) - \xi e^{2}A_0^{2}v_F^{2}\sigma_{z}$. In the limit when the driving frequency $\omega$ is large compared to the other energy scales, a proper description of the system is the effective time-independent Hamiltonian\cite{Kitagawa11,Goldman14,Grushin14,bukov15}, which reads

\begin{eqnarray}
\mathcal{H}_{\rm eff}(k)&=&\mathcal{H}_{0}+\sum_{n\geq1}\frac{[\mathcal{H}_{+n},\mathcal{H}_{-n}]}{n\omega}+\mathcal{O}(\frac{1}{\omega^{2}})
\nonumber\\
&=&v_F(k_{x}\sigma_{x}+k_{y}\sigma_{y}) + [m_z-t_z + \frac{1}{2}t_z k_z^2\nonumber\\ 
&+&\frac{m_0}{2}(k_x^2+ k_y^2)]\sigma_{z}+\frac{2}{\omega}\xi e^{2}A_0^{2}v_{F}m_0(k_{x}\sigma_{x} + k_{y}\sigma_{y})\nonumber\\
&-& \frac{1}{\omega}\xi e^{2}A_0^{2}v_F^{2}\sigma_{z}\nonumber\\
&=&v_{eff}(k_{x}\sigma_{x}+k_{y}\sigma_{y}) +[m'_z -t_z +\frac{1}{2}t_z k_z^2 \nonumber\\
&+& m_0(k_x^2+k_y^2)]\sigma_{z}
\label{effham1}
\end{eqnarray}

\noindent where $v_{eff}=v_{F}[1+\xi \frac{2}{\omega}m_0 (e A_0)^2] $ and $ m'_{z}=m_{z}-\frac{1}{\omega}\xi (e A_0)^2 v_F^{2} $. The mass term $ m_z $ get renormalized due to first order photon processes in single WSM. Thus, the effect of left circularly polarized light($ \xi=-1 $) has same NI phase with larger band gap in comparison to its undressed counterpart. Therefore, we will focus our study only to right circularly polarized light($ \xi=1 $) since this give rise to Floquet single-WSMs phase for $m'_z < t_z$. The terms proportional to $ \sigma_{x,y} $ serve to renormalize the velocities in the x,y directions at $ \Gamma $ point in the Brillouin zone. For $\mid m'_z\mid/t_z<1 $, there appear two Weyl points at $ \textbf{k}_c=(0,0,\pm b ) $, with $b=\sqrt{m_2/m_1}$, $m_2=t_z-m'_z$ and $m_1=0.5 t_z$ at which the valence and conduction bands touch at zero energy. Then the Hamiltonian Eq.(\ref{effham1}) for J=1 reduces to a form of WSMs,

\begin{equation}
\mathcal{H}_{1,\rm eff}(k)= v_{eff}(k_{x}\sigma_{x}+k_{y}\sigma_{y}) \pm 2\sqrt{m_1 m_2}(k_z -b)\sigma_z \label{wsm1}
\end{equation}

The coefficient $2\sqrt{m_1 m_2}$ act as the effective velocity $v_z$ along z-direction. The energy spectra of $\mathcal{H}_{\rm
eff}$ are given by

\begin{eqnarray}
\tilde{E}_{\pm, k}=\pm
\sqrt{v_{eff}^{2}(k_x^2+k_y^2)+4m_1m_2(k_{z}-b)^{2}}. \label{5}
\end{eqnarray}

\begin{figure}[h]       
\fbox{\includegraphics[scale=.4]{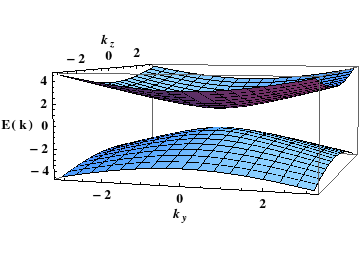}}   
\hspace{30px}
\fbox{\includegraphics[scale=.4]{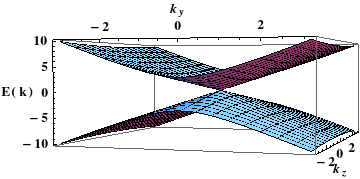}}
\hspace{30px}
\fbox{\includegraphics[scale=.22]{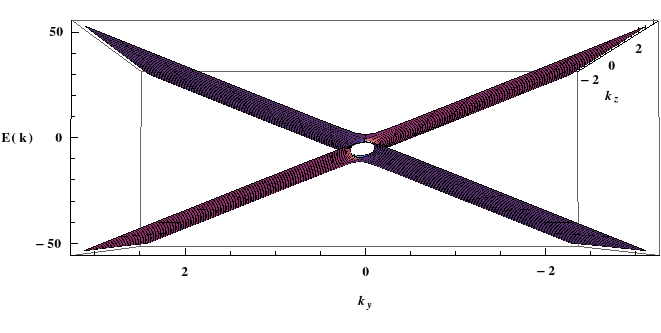}}
\caption{Evolution of Floquet energy spectrum for the single-WSM under increasing amplitude of laser light, $A_0 =0.0, 1.4, 2 $ from left to right. The normal insulating gap decreases with increasing $A_0$ which becomes zero for $A_0 \approx 1.4$. The gap reopens with further increase of $A_0$ with inverted order along with a pair of Weyl points located symmetrically along $k_y$. Here we set $t_z =2.0$, $m_z =3.0$, $v_F=1.0$, $m_0=1.0$, $\omega =2.0$}
\label{figswsm}
\end{figure}

The evolution of energy spectrum is shown in Fig.(\ref{figswsm}) where WSM phase arises with two Weyl points with increasing of the intensity of radiation. The separation between the Weyl points can be tuned by laser light.

\subsection{Double-WSMs}
The non-interacting low energy effective Hamiltonian for a single double-Weyl fermion system near $\Gamma$ point can be written as~\cite{Jian15, ahna16} 

\begin{equation}
 \mathcal{H}=\frac{k_{x}^{2}-k_{y}^{2}}{2m}\sigma_{x}+\frac{k_{x}k_{y}}{m}\sigma_{y}+M(\textbf{k})\sigma_{z},
\end{equation}

\noindent where $M(\textbf{k}) \approx m_z-t_z + \frac{1}{2}t_z  k_z^2 + \frac{m_0}{2}(k_x^2+k_y^2)$, $m$ is the effective mass along the x-y plane. $\sigma_{x}$. For $m_z > t_z$ , the mass term $M_z$ remains positive throughout the Brillouin zone, thus the system is trivial insulator. The gap parameter $m_z$ can be tuned by laser light  which results in WSMs i.e. $m_z < t_z$ .\\

Following the same strategy as in the case of single-WSMs case, we can expand the time-periodic Hamiltonian for double-WSMs as
$\mathcal{H}(t,k)=\sum_{n}\mathcal{H}_{n}(k)e^{in\omega t}$ with
\begin{eqnarray}
\mathcal{H}_{0}(p)&=&\frac{k_{x}^{2}-k_{y}^{2}}{2m}\sigma_{x}+\frac{k_{x}k_{y}}{m}\sigma_{y}+ M(\textbf{k}) \sigma_{z},\nonumber\\
\mathcal{H}_{\pm1}(k)&=& \frac{eA_{0}}{m}(k_{x}\pm i \xi k_{y})(\sigma_{x}\mp i \xi \sigma_{y})\nonumber\\
&+& m_0 e A_0 (k_{x}+ i \xi k_{y})\sigma_z,\nonumber\\
\mathcal{H}_{\pm2}(k)&=&\frac{e^{2}A_{0}^{2}}{2 m}(\sigma_x \mp i \sigma_y),
\end{eqnarray}
and $\mathcal{H}_{n}=0$ for $|n|>2$. Using the commutation relations for the Pauli matrices, we obtain the commutator in this case as $[H_{1},H_{-1}]= -4 \xi \frac{e^{2}A_0^{2}}{m^2}(k_x^2+k_y^2)\sigma_{z} + 8\xi m_{0}(e A_0)^2[\frac{k_{x}^{2}-k_{y}^{2}}{2m}\sigma_{x}+\frac{k_{x}k_{y}}{m}\sigma_{y}]$ and $[H_{2},H_{-2}]= -\xi\frac{e^{4}A_0^{4}}{m^2}\sigma_{z}$. Therefore, the effective time-independent Hamiltonian for double WSM 

\begin{eqnarray}
\mathcal{H}_{\rm eff}(k)&=&\mathcal{H}_{0}+\sum_{n\geq1}\frac{[\mathcal{H}_{+n},\mathcal{H}_{-n}]}{n\omega}+\mathcal{O}(\frac{1}{\omega^{2}})
\nonumber\\
&=&\frac{k_{x}^{2}-k_{y}^{2}}{2m}\sigma_{x}+\frac{k_{x}k_{y}}{m}\sigma_{y}+[m_z-t_z + \frac{1}{2}t_z k_z^2\nonumber\\ 
&+& \frac{m_0}{2}(k_x^2+k_y^2)]\sigma_{z}-\frac{4\xi}{\omega} \frac{e^{2}A_0^{2}}{m^2}(k_x^2+k_y^2)\sigma_{z}+\frac{8}{\omega}\xi \nonumber\\
&& m_{0}(e A_0)^2[\frac{k_{x}^{2}-k_{y}^{2}}{2m}\sigma_{x}
+\frac{k_{x}k_{y}}{m}\sigma_{y}]-\frac{\xi}{2 \omega}\frac{e^{4}A_0^{4}}{m^2}\sigma_{z}\nonumber\\
&=&\frac{k_{x}^{2}-k_{y}^{2}}{2m'}\sigma_{x}+\frac{k_{x}k_{y}}{m'}\sigma_{y}+[m'_z -t_z +\frac{1}{2}t_z k_z^2 \nonumber\\
&+& \frac{m'_0}{2}(k_x^2+k_y^2)]\sigma_{z}
\label{effham}
\end{eqnarray}

\noindent where $ m'=\frac{m}{[1+\frac{8}{\omega}\xi m_{0}(e A_0)^2]} $, $ m'_{z}=m_{z}-\frac{\xi}{2 \omega}\frac{e^{4}A_0^{4}}{m^2} $ and $ m'_{0}=m_{0}-\frac{8\xi}{\omega} \frac{e^{2}A_0^{2}}{m^2} $. the mass term $ m_z $ get renormalized due to second order photon processes in double WSM.The laser light renormalize the parameter $ m'_{0} $  close to zero at a particular frequency for a fixed strength of light which is necessary for the WSM phase to occur.\\

We are interested only in right circularly polarized light($ \xi=1 $) since this give rise to Floquet double-WSMs phase. The left circularly polarized light($ \xi=-1 $) has only NI phase with larger band gap in comparison with its undressed part. For $\mid m'_z\mid/t_z<1 $, there appear two Weyl points at $ \textbf{k}_c=(0,0,\pm b ) $, with $b=\sqrt{m_2/m_1}$, $m_2=t_z-m_z$ and $m_1=0.5 t_z$ at which the valence and conduction bands touch at zero energy. Then the Hamiltonian Eq.(\ref{effham}) for J=2 reduces to a form of WSMs,

\begin{equation}
\mathcal{H}_{2,\rm eff}(k)=\frac{k_{x}^{2}-k_{y}^{2}}{2m'}\sigma_{x}+\frac{k_{x}k_{y}}{m'}\sigma_{y} \pm 2\sqrt{m_1 m_2}(k_z -b)\sigma_z
\end{equation}


The energy spectra of $\mathcal{H}_{\rm
eff}$ are given by
\begin{eqnarray}
\tilde{E}_{\pm, k}=\pm
\sqrt{\Bigl[\frac{k_{x}^{2}-k_{y}^{2}}{2m'}\Bigr]^{2}+\Bigl[\frac{k_{x}k_{y}}{m'}\Bigr]^{2}+4m_1m_2(k_{z}-b)^{2}}. \label{5}\nonumber\\
\end{eqnarray}

\begin{figure}[h]       
    \fbox{\includegraphics[scale=.425]{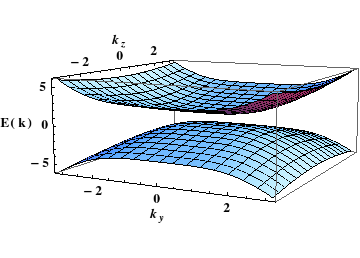}}   
    \hspace{30px}
    \fbox{\includegraphics[scale=.27]{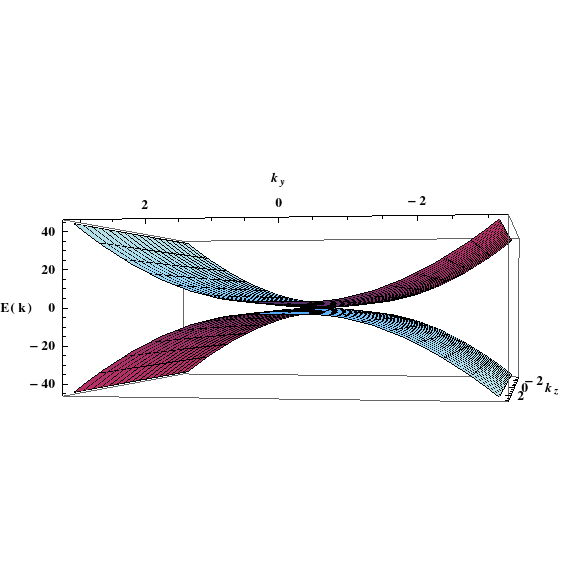}}
    \hspace{30px}
    \fbox{\includegraphics[scale=.22]{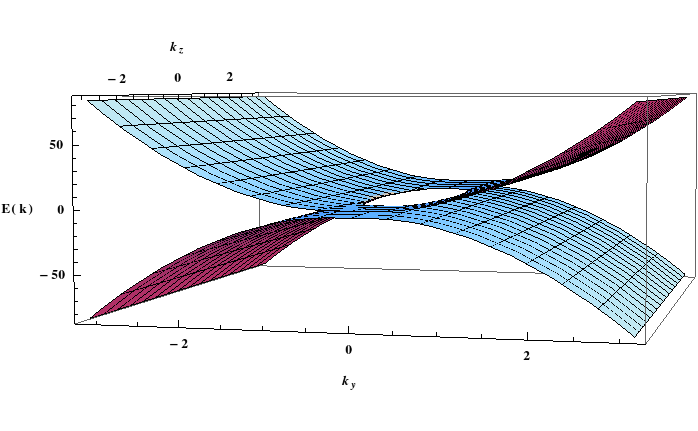}}
\caption{Evolution of Floquet energy spectrum for the double-WSM under increasing amplitude of laser light, $A_0 =0.0, 1.4, 2$ from left to right. The normal insulating gap decreases with increasing $A_0$ which becomes zero for $A_0 \approx 1.4$. The gap reopens with further increase of $A_0$ with inverted order along with a pair of Weyl points located symmetrically along $k_y$. Here we set $t_z =2.0$, $m_z =3.0$, $m=1.0$, $m_0=1.0$, $\omega =2.0$}
\label{figdwsm}
\end{figure}
The evolution of energy spectrum is shown in Fig.(\ref{figdwsm}) where double-WSM phase arises with two Weyl points are separated along z-direction with increasing the intensity of laser light. The separation between the Weyl points can be controlled by laser light.

\subsection{Triple-WSMs}
\label{sec:Floquet}
The low-energy Hamiltonian Eq. (\ref{twsm}) for a triple-WSMs near $k = (0, 0, 0)$ can be written as~\cite{ahna16}

\begin{equation}
 \mathcal{H}=\alpha_{3} k_{x}(k_{x}^{2}-3k_{y}^{2})\sigma_{x}+ \alpha_{3}k_{y}(k_{y}^{2}-3k_{x}^{2})\sigma_{y}
+M'(\textbf{k})\sigma_{z}
\end{equation}

\noindent where $M(\textbf{k}) \approx m_z-t_z + \frac{1}{2}t_{z} k_z^2 + \frac{m_0}{2}(k_x^2+k_y^2)$.

Following the same strategy as in previous cases, we can expand the time-periodic Hamiltonian for triple-WSMs as
$\mathcal{H}(t,k)=\sum_{n}\mathcal{H}_{n}(k)e^{in\omega t}$ with

\begin{eqnarray}
\mathcal{H}_{0}(k)&=& \alpha_{3} k_{x}(k_{x}^{2}-3k_{y}^{2})\sigma_{x}+ \alpha_{3}k_{y}(k_{y}^{2}-3k_{x}^{2})\sigma_{y}\nonumber\\
&+& M'_{z}\sigma_{z},\nonumber\\
\mathcal{H}_{\pm1}(k)&=& \frac{3\alpha_3 eA_{0}}{2}(k_x \pm i\xi k_y)^2(\sigma_{x}\pm i\xi \sigma_y)\nonumber\\
&+& m_0eA_{0}(k_{x}\mp i k_y)\sigma_{z},\nonumber\\
\mathcal{H}_{\pm2}(k)&=&\frac{3 \alpha_3}{2}e^{2}A_{0}^{2}(k_x \pm i \xi k_y)(\sigma_{x}\pm i\xi \sigma_y),\nonumber\\
\mathcal{H}_{\pm 3}(k)&=&\frac{1}{2} \alpha_3 e^{3}A_{0}^{3}(\sigma_{x}\pm i\xi \sigma_y),
\end{eqnarray}

\noindent and $\mathcal{H}_{n}=0$ for $|n|>3$. Using the commutation relations for the Pauli matrices, we obtain the commutator in this case as $[H_{1},H_{-1}]= 9\xi\alpha_3^2e^{2}A_0^{2}
(k_x^2+k_y^2)^2\sigma_{z}-6\xi e^{2}A_0^{2}m_0 [ \alpha_{3} k_{x}(k_{x}^{2}-3k_{y}^{2})\sigma_{x}+ \alpha_{3}k_{y}(k_{y}^{2}-3k_{x}^{2})\sigma_{y}]$, $[H_{2},H_{-2}]= 9\xi \alpha_3^2(e A_0)^4(k_x^2+k_y^2)\sigma_z$ and $[H_{3},H_{-3}]= \xi \alpha_3^2(e A_0)^6\sigma_z$. Thus, the effective time-independent Hamiltonian for triple-WSM 

\begin{eqnarray}
\mathcal{H}_{\rm eff}(k)&=& \alpha_{3} k_{x}(k_{x}^{2}-3k_{y}^{2})\sigma_{x}+ \alpha_{3}k_{y}(k_{y}^{2}-3k_{x}^{2})\sigma_{y}+[ m_z\nonumber\\
&-&t_z + \frac{1}{2}t_{z} k_z^2 +\frac{m_0}{2}(k_x^2+k_y^2)]\sigma_{z}\nonumber\\
&+&{\frac{1}{\omega}}\{ 9\xi\alpha_3^2e^{2}A_0^{2}(k_x^2+k_y^2)^2\sigma_{z}
-6\xi e^{2}A_0^{2}m_0\nonumber\\
&&[ \alpha_{3} k_{x}(k_{x}^{2}-3k_{y}^{2})\sigma_{x}+\alpha_{3}k_{y}(k_{y}^{2}-3k_{x}^{2})\sigma_{y}]\}\nonumber\\
&+& \frac{9}{2\omega}\xi \alpha_3^2(e A_0)^4(k_x^2+k_y^2)\sigma_z + \frac{1}{3 \omega}\xi \alpha_3^2(e A_0)^6\sigma_z \nonumber\\
&=& \alpha'_{3} k_{x}(k_{x}^{2}-3k_{y}^{2})\sigma_{x}+ \alpha'_{3}k_{y}(k_{y}^{2}-3k_{x}^{2})\sigma_{y}\nonumber\\
&+&[ m'_z-t_z + \frac{1}{2}t_{z} k_z^2 +m'_0(k_x^2+k_y^2)\nonumber\\
&+&m''_0(k_x^2+k_y^2)^2]\sigma_{z} \label{effham_twsm}
\end{eqnarray}

\noindent where $\alpha'_{3} =\alpha_{3}[1-{\frac{6}{\omega}}\xi e^{2}A_0^{2}m_0]$, $ m'_z=m_z[1+\frac{1}{3 \omega}\xi \alpha_3^2(e A_0)^6] $ and $ m'_0=m_0+\frac{9}{\omega}\xi \alpha_3^2(e A_0)^4$. The mass term $ m_z $ get renormalized due to third order photon processes in triple WSM. The laser light renormalize the parameter $ m'_{0} $ close to zero which is necessary for the WSM phase to occur. We have neglected the fourth powers of momentum term due to its small value near BZ center.\\

We are interested only left-circularly polarized light($ \xi=-1 $) since this give rise to Floquet triple-WSMs phase. The right circularly polarized light($ \xi=1 $) has only NI phase with larger band gap in comparison with undressed part. For $\mid m'_z\mid/t_z<1 $, there appear two Weyl points at $ \textbf{k}_c=(0,0,\pm b ) $, with $b=\sqrt{m_2/m_1}$, $m_2=t_z-m_z$ and $m_1=0.5 t_z$ at which the valence and conduction bands touch at zero energy. Then the Hamiltonian Eq.(\ref{effham_twsm}) for J=3 reduces to a form of WSMs,

\begin{eqnarray}
\mathcal{H}_{3, \rm eff}(k)&=&\alpha'_{3} k_{x}(k_{x}^{2}-3k_{y}^{2})\sigma_{x}+ \alpha'_{3}k_{y}(k_{y}^{2}-3k_{x}^{2})\sigma_{y}\nonumber\\
&&\pm 2\sqrt{m_1 m_2}(k_z -b)\sigma_z
\end{eqnarray}


The energy spectra of $\mathcal{H}_{\rm
eff}$ are given by

\begin{equation}
\tilde{E}_{\pm, k}=\pm\sqrt{\hspace{-.35cm}\begin{aligned}
&&\Bigl[\alpha'_{3} k_{x}(k_{x}^{2}-3k_{y}^{2})\Bigr]^{2}+\Bigl[\alpha'_{3}k_{y}(k_{y}^{2}-3k_{x}^{2})\Bigr]^{2}\\
&&+4m_1m_2(k_{z}-b)^{2}
\end{aligned}}
\end{equation}

\begin{figure}[h]       
\fbox{\includegraphics[scale=.24]{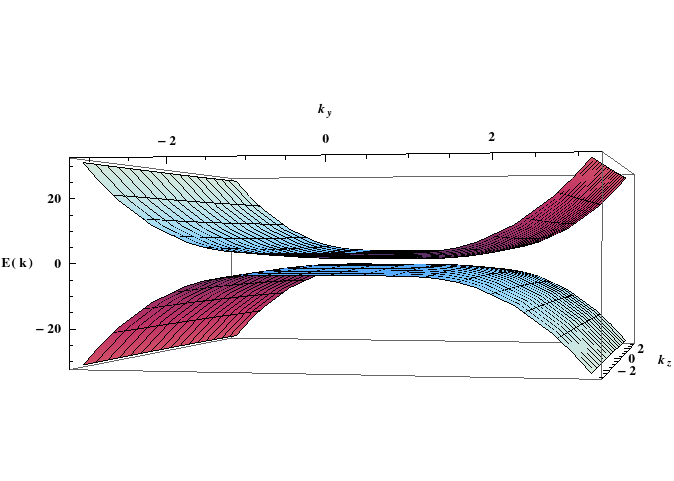}}   
\hspace{30px}
\fbox{\includegraphics[scale=.161]{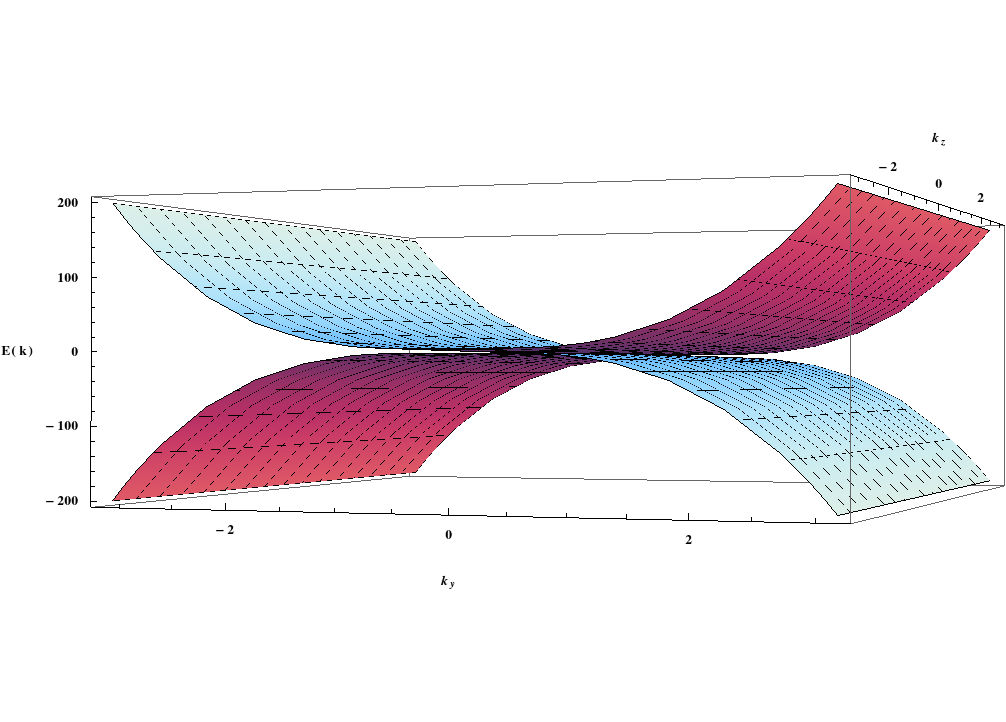}}
\hspace{30px}
\fbox{\includegraphics[scale=.215]{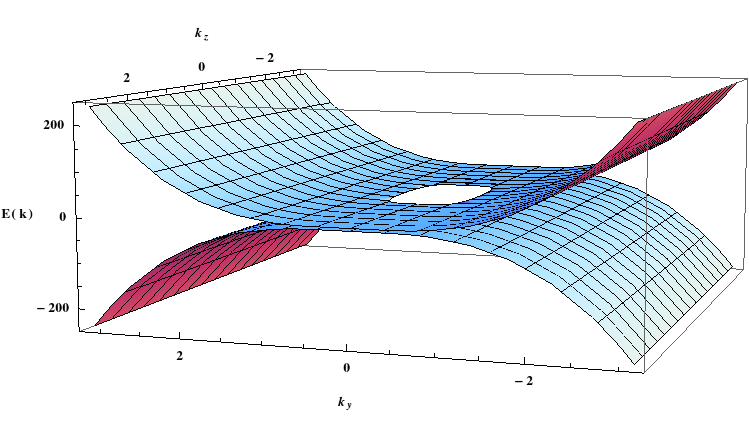}}
\caption{Evolution of Floquet energy spectrum for the double-WSM under increasing amplitude of laser light, $A_0 =0.0, 1.4, 2$ from left to right. The normal insulating gap decreases with increasing $A_0$ which becomes zero for $A_0 \approx 1.4$. The gap reopens with further increase of $A_0$ with inverted order along with a pair of Weyl points located symmetrically along $k_y$. Here we set $t_z =2.0$, $m_z =3.0$, $\alpha_3=1.0$, $m_0=1.0$, $\omega =2.0$}
\label{figtwsm}
\end{figure}

The evolution of energy spectrum is shown in Fig.(\ref{figdwsm}) where triple-WSM phase arises with two Weyl points are separated along z-direction with increasing the intensity of laser light. The separation between the Weyl points can be increased or decreased by laser light.

\section{Anomalous Hall Conductivity}
The main consequences of the topological phase transition from NI phase to WSM phase is the emergence of anomalous Hall conductivity. Within the linear response theory \cite{Oka09,Yan2016}, the formula for the Hall conductivity $\sigma_{xy}$ for the laser light beams along the $z$ direction is given by

\begin{eqnarray}
\sigma_{xy}=e^{2}\int\frac{d^{3}k}{(2\pi)^{3}}
\sum_{\alpha}f_{\alpha}(k)[\nabla_{\bf k}\times\mathcal{A}_{\alpha}(\textbf{k})]_z,\label{S1}
\end{eqnarray}

\noindent where $\alpha\equiv (i,m)$, where $i (= \pm)$ labels the original band index and $n$ labels the Floquet
index\cite{Oka09}; $\mathcal{A}_{\alpha}(\bf k)$ is the Berry connection, and $f_\alpha$ is the fermionic occupation function. Thus, the Hall conductivity depends not only on the Berry curvature but also on the fermion occupation which depends on both indices $(i,m)$. For the case when the system is close to equilibrium,i.e., $f_{i,n}(k)=\delta_{n,0}f_{FD}(E_{i}(\bf k))$, where $f_{FD}(E_{i}(\textbf{k}))=1/(e^{(E_{i}(\textbf{k})-\mu)/T}+1)$ is the Fermi-Dirac distribution. Neglecting second and higher order terms in  $\mathcal{O}(1/\omega^2)$, we get expression for the dc  Hall conductivity,

\begin{eqnarray}
\sigma_{xy}(A_0, T,\mu)
&=&\frac{e^{2}}{2}\int\frac{d^{3}k}{(2\pi)^{3}}[\hat{\bf d}\cdot(\frac{\partial
\hat{\bf d}}{\partial k_{y}}\times\frac{\partial \hat{\bf d}}{\partial k_{z}})]\nonumber\\
&&[f_{FD}(E_{+}(\textbf{k}))-f_{FD}(E_{-}(\textbf{k}))],
\label{equilibrium}
\end{eqnarray}

\noindent where $\hat{\bf d}=\bf d/\mid d \mid$ and $\textbf{d}=\mathrm{Tr}[\sigma \mathcal{H}_{J,\rm eff}]$ . For finite temperature ($T\neq0$) or finite chemical potential system ($\mu\neq0$), analytic simplification of Eq.(\ref{equilibrium}) is not possible, and we need to treat Eq.(\ref{equilibrium}) numerically. Taking $\mu=T=0$ in Eq.(\ref{equilibrium}) leads to

\begin{eqnarray}
\sigma_{xy}(A_0)&=&\frac{e^{2}}{2}\int_{-k_c}^{k_c}\frac{dk_z}{2\pi^2}C(k_z),
\end{eqnarray}

\noindent Where $C(k_z) $ is the topological(Chern) invariant defined in the momentum space as

\begin{equation}
C(k_z)=\frac{1}{{4 \pi}}\int dk_x dk_y
\hat{\bf d}\cdot(\frac{\partial
\hat{\bf d}}{\partial k_{x}}\times\frac{\partial \hat{\bf d}}{\partial k_{y}})
\end{equation}

We can therefore evaluate its Chern number C for each $k_z$ fixed plane. We find that $C=J$ for the planes with $-k_c < k_z<k_c$ otherwise $C=0$. Let us consider the simplest case of multi-WSM in which the two Weyl points with opposite chirality are located at $ k_z=k_c=\pm b\hat{z} $, where $b=\sqrt{m_2/m_1}=\sqrt{(t_z-m'_z)/0.5 t_z}=\sqrt{(t_z-m_z+\frac{A_0^{2J}\alpha_J^2}{J\omega})/0.5 t_z}$. Therefore, the expression for Hall conductivity simplify to

\begin{eqnarray}
\sigma_{xy}(A_0)&= &J\frac{e^{2}}{h \pi}b=J\frac{e^{2}}{h \pi}\sqrt{(t_z-m_z+\frac{A_0^{2J}\alpha_J^2}{J\omega})/0.5 t_z}\nonumber\\
\end{eqnarray}

\noindent where we have restored the Planck constant $h$. It is readily seen that $\sigma_{xy}(A_0)$ is proportional to the distance between the two Weyl points which can be easily tuned by intensity of the incident laser light. \\

\section{Conclusion and Summary}
In conclusion, we summarize the main findings of the present manuscript. We have investigated a time periodic dynamics of multi-WSMs under off-resonant light. We have seen the effect of circularly polarized light in normal insulating phase which can be tunable to Weyl semimetal phase by increasing the intensity of laser light. Additionally, the positions of Weyl nodes can be tuned to lasee light. The emergence of WSM phase is accomplished with large anomalous Hall conductivity and are proportional to $A_0^{2J}$ for a fixed frequency $ \omega $. Our predictions could be measured in transport measurements or pump-probe angle resolved photoemission spectroscopy.
\label{}
 
\section{Acknowledgements}
We acknowledge helpful discussions with N. S. Vidhyadhiraja. This work is supported by Science and Engineering Research Board (SERB), India for the SERB National Post doctoral Fellowship. I would like to thank JNCASR for the facilities for this work.
  



\begin{thebibliography}{00}
\bibitem{argument1}
In this paper, we have assumed that a laser beam is incident in the z-direction. The multi WSMs Hamiltonians are anisotropic in x and y-z directions for $ J>1 $. In this case only, the laser field induce Weyl points in the systems.
\end{thebibliography}


\end{document}